\documentclass[pra,twocolumn,bibnotes,superscriptaddress]{revtex4-1}
\usepackage{graphicx}
\usepackage{soul} 
\usepackage{amsmath}
\usepackage{color}

\begin{document}
	
\title{Observation of a non-equilibrium steady state of cold atoms in a moving optical lattice}
	
\author{Kyeong Ock Chong}
\author{Jung-Ryul Kim}
\author{Jinuk Kim}
\author{Seokchan Yoon}
\altaffiliation[Current Address: ]{IBS Center for Molecular Spectroscopy and Dynamics, Korea University, Seoul 136-701, Korea}
\author{Sungsam Kang}
\altaffiliation[Current Address: ]{Laser Biomedical Research Center, G.~R.~Harrison Spectroscopy Laboratory, Massachusetts Institute of Technology, Cambridge, MA 02139, U.S.A.}
\author{Kyungwon An}
\email[E-mail: ]{kwan@phys.snu.ac.kr} 
\affiliation{Department of Physics and Astronomy, Institute of Applied Physics, Seoul National University, Seoul 08826, Korea}
	
\begin{abstract}
We investigated non-equilibrium atomic dynamics in a moving optical lattice via observation of atomic resonance fluorescence spectrum. 
A three-dimensional optical lattice was generated in a phase-stabilized magneto-optical trap (MOT) and the lattice was made to move by introducing a detuning between the counter-propagating trap lasers. 
A non-equilibrium steady states (NESS's) of atoms was then established in the hybrid of the moving optical lattice and the surrounding MOT. 
A part of atoms were localized and transported in the moving optical lattice and the rest were not localized in the lattice while trapped as a cold gas in the MOT.
These motional states coexisted with continuous transition between them. 
As the speed of the lattice increased, the population of the non-localized state increased in a stepwise fashion due to the existence of bound states at the local minima of the lattice potential. 
A deterministic rate-equation model for atomic populations in those motional states was introduced in order to explain the experimental results. 
The model calculations then well reproduced the key features of the experimental observations, confirming the existence of an NESS in the cold atom system.
\end{abstract}
\maketitle	
	
	
A collection of cold atoms in a periodic optical potential\cite{Guidoni1999} is a useful tool for metrological applications as well as fundamental studies. 	
Atomic clocks using optical lattices are reported to have ultrahigh resolution\cite{Bloom2014}. 
Studies on nonlinear dynamics and chaotic motion due to photons recoil and lattice amplitude modulation\cite{PhysRevA.78.043413,PhysRevE.89.012917,Hensinger2001} have been proposed based on the atomic motion corresponding to classically-well-understood harmonic oscillators in the optical lattice.	
Landau-Zener tunneling\cite{PhysRevLett.99.190405} by lattice acceleration and Mott insulator-metal transition\cite{Jordens2008} by using periodic modulation of the lattice amplitude have been investigated. 
As seen in these examples, cold atom systems have experimental advantages in realizing ideal lattice models. They are much more controllable than real electron systems\cite{RevModPhys.86.779}. 

Recently, non-equilibrium dynamics of cold atoms has become an active research topic\cite{MOON20171,Bloch2012}, expanding our understanding based on the existing equilibrium physics.
In particular, non-equilibrium steady states (NESS's), the stationary states with continuous flows among them, have drawn much interest among various non-equilibrium phenomena. 
For example, biochemical reactions such as ATP hydrolysis and other biomolecular motors are described by a stochastic NESS theory\cite{ZHANG20121}. In quantum physics, NESS is suggested as a possible pathway to entanglement at high temperature\cite{Hsiang2015} and generation of high-temperature Bose-Einstein condensates\cite{PhysRevLett.119.140602}.
Associated with an array of particles, a particular interest lies in
asymmetric simple exclusion process (ASEP)\cite{PhysRevLett.93.238102}, describing a driven diffusive system of one-dimensional lattice of particles hopping to other sites at a certain rate along the lattice. 
A model of ASEP with the Langmuir kinetics, which includes absorption and desorption of particles in a lattice connected to a reservoir, is suggested\cite{PhysRevE.70.046101}
to describe the NESS system of mRNA translation\cite{0034-4885-74-11-116601}, for example. 
Related with cold atom systems, several theoretical investigation on NESS\cite{PhysRevB.88.245114,PhysRevB.81.144301,PhysRevLett.111.240405} as well as experimental observation of bistability in a driven superfluid have been recently reported\cite{PhysRevLett.116.235302}.
However, clear experimental observation of NESS, also with ASEP characteristics, in a cold atom system has not been reported yet. 
	
A hybrid trap combining a magneto-optical trap (MOT) and an optical lattice potential is a possible approach to facilitates such studies.
This hybrid trap can be generated by a passively-phase-stabilized MOT\cite{Rauschenbeutel199845} of which time-dependent phase fluctuation is canceled out to produce a stable optical lattice. 
The atoms affected by sub-Doppler cooling are then strongly confined at the local minima of the lattice potential\cite{Westbrook1997}. They are in the so-called Lamb-Dicke regime(LDR)\cite{PhysRev.89.472} and their spectrum exhibits a Rayleigh peak as well as Raman sidebands spaced by the vibrational frequency $\Delta\omega_{\rm osc}$ of the optical lattice.
Previously, matter-wave tunneling among the optical lattice sites was observed by measuring the resonance fluorescence spectrum of the trapped atoms\cite{wrkim2011}.
	
It is suggested that 
the atoms interacting with radiation fields under an external drive 
can be in a NESS\cite{10.1038/ncomms7977}. 
Likewise, in the aforementioned hybrid trap, a non-equilibrium situation can be achieved if the optical lattice is made to move at a certain speed. 
In this case, the lattice motion introduces a potential modulation in time and thus acts as an external drive. 
Moreover, the photon recoils present in the system make the atomic motion stochastic. 
Because of these processes, two different atomic motional states,
one localized in the optical lattice and the other trapped as a cold gas in the background MOT,
can coexist in the steady state with continuous flows between them.  
The coexistence of such motional states in the steady state would then establish a NESS.

\begin{figure*}
\includegraphics[width=\textwidth]{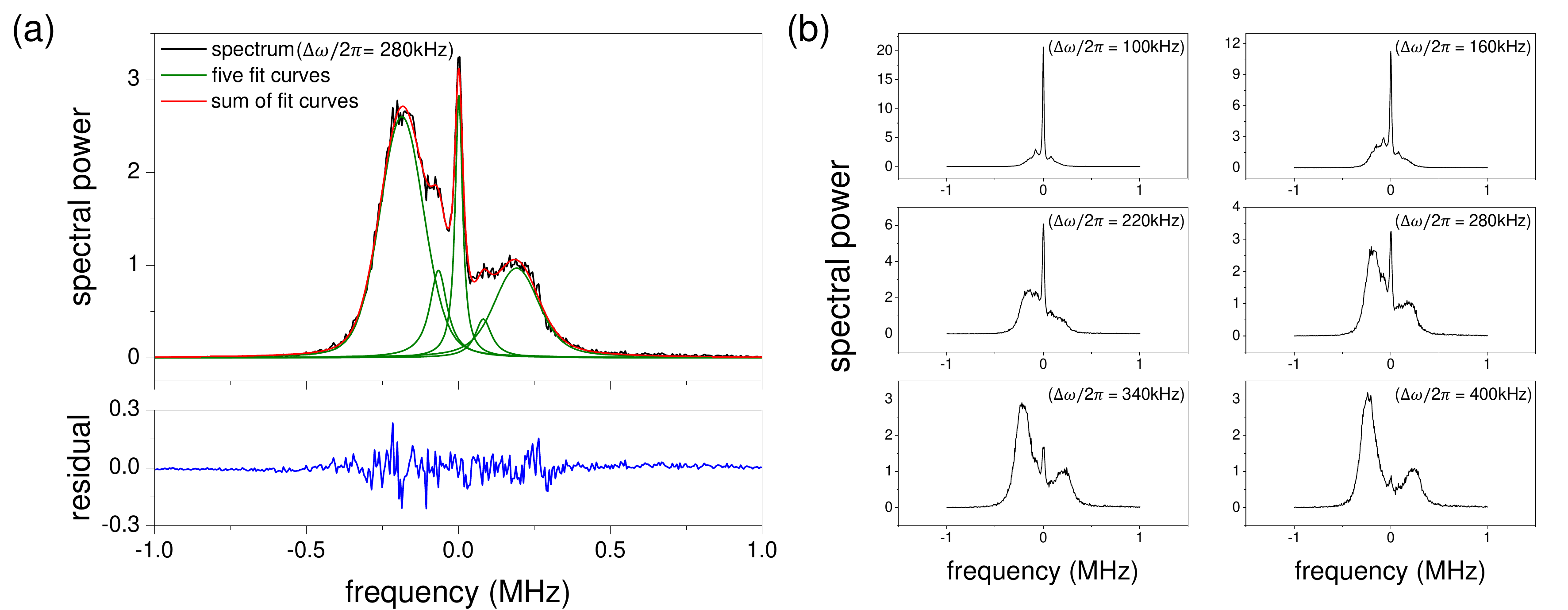}
\caption{{\bf Observed resonance fluorescence spectrum of atoms.} (a) Fitting the fluorescence spectrum. The broad peaks at both ends represent the atoms nonlocalized in the lattice but trapped in the background MOT. The narrow three peaks in the center represent the atoms localized in the lattice potential minima. The zero frequency corresponds to the atomic transition frequency minus 3$\Gamma$ (red detuning). In the present frequency scale, the frequency of one trap laser is at +140kHz and that of the other is at -140kHz ($\Delta \omega/2\pi$=280kHz).	
(b) Fluorescence spectrum at various lattice speeds. The population of nonlocalized(localized) atoms increase(decrease) as the speed of the lattice increases.
}
\label{fig1}
\end{figure*}
In this paper, we investigated the dynamics of cold atoms in an NESS with ASEP characteristics in a hybrid trap of an moving optical lattice and a background MOT.
The optical lattice was made to move by introducing a frequency detuning between the counter-propagating trap lasers in a phase-stabilized MOT. 
Dynamics of the atoms were observed non-destructively by measuring the resonance fluorescence spectrum of atoms with a photon-counting-based heterodyne technique\cite{Hong:06} for various lattice speeds. 
The high-resolution fluorescence spectrum then revealed different atomic motional states. 		
The observed atomic dynamics were analyzed with a deterministic 
rate equation model describing transitions among the vibrational states of the optical lattice and the MOT cold-gas state. 
As a result, we could show that the atoms are in an NESS. 
Atoms localized by the lattice potential can be nonlocalized via tunneling through the lattice potential barrier due to the lattice motion. On the other hand, the nonlocalized atoms can be localized again in the lattice via momentum diffusion due to photon recoils\cite{PhysRevLett.107.243002} in the background MOT. These transitions occur simultaneously and continuously while the population of each motional states are stationary.	\\

\noindent
{\bf \large Results}
	
In our experiment, the details of which are described in Methods, the resonance fluorescence spectrum of atoms was measured for various trap-laser frequency difference $\Delta\omega$ between 0 and 500kHz, by which the speed of the moving optical lattice was determined to be between 0 and 0.34m/s.
Five spectral peaks were observed in the spectrum when optical lattice was moving. We analyzed the spectra by fitting them with appropriate profiles as shown in Fig.\ \ref{fig1}(a). 
The two broad peaks at both ends are made by the atoms that are not localized in the lattice but trapped as a cold gas in the MOT. 
We will call them `nonlocalized' in the optical lattice in short in the discussions below.
These atoms scatter off the two different-frequency trap lasers, creating a peak around each trap laser frequency. 
These atoms are also continuously subject to the sub-Doppler cooling by polarization gradient, which results in a non-Gaussian velocity distribution\cite{PhysRevA.54.2275}. 
The distribution is well fit by the Voigt profile.
The middle three picks correspond to the atoms that are localized at the potential minima of the optical lattice and thus move along with the lattice. 
Let us call them `localized' in the optical lattice in short from now on. 
These atoms are in a state associated with the Lamb-Dicke regime (LDR). The atoms in LDR, confined in the optical lattice, exhibit three Lorentzian peaks: a Rayleigh peak in the center and Raman sidebands at both sides. 
\begin{figure*}		
\includegraphics[width=0.939\textwidth]{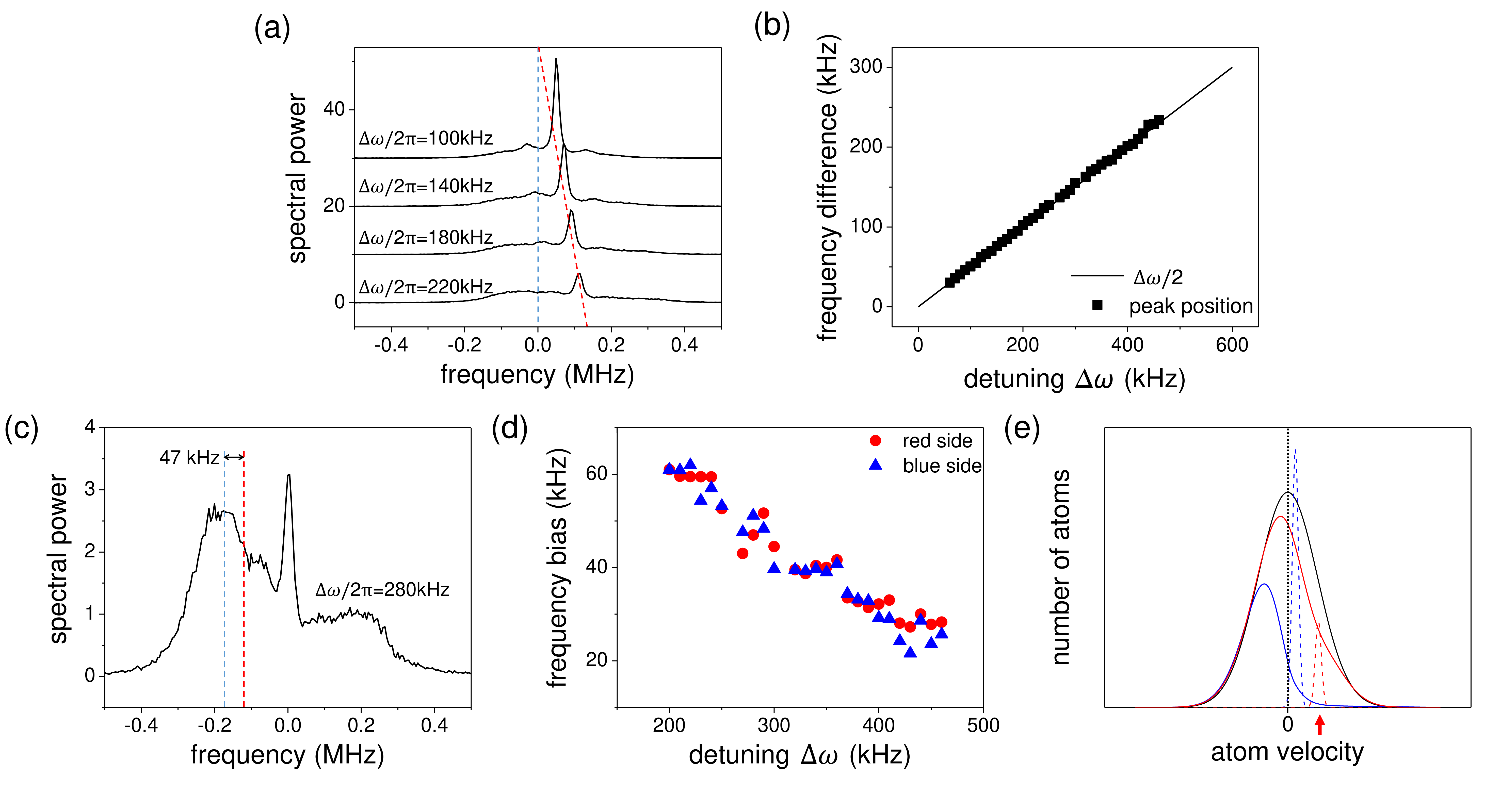}
\caption{{\bf Evolution of spectral peaks with the trap-laser frequency difference.} (a) Evolution of the position of the Rayleigh peaks as the speed of the lattice increases. The frequency is measured with respect to the lower frequency $\omega$ of the trap lasers.
(b) Rayleigh peaks occur at frequency $\omega+\Delta\omega/2$, indicating that the contributing atoms move along with the lattice, {\em i.e.}, localized at the potential minima of the lattice. (c) The center frequencies of the peaks corresponding to the nonlocalized atoms are slightly shifted away from the trap laser frequencies. It is because the momentum distribution of the nonlocalized atoms are biased opposite to the lattice moving direction. The frequency is measured with respect to $\omega+\Delta\omega/2$ and thus the trap laser frequencies are at $\pm\Delta\omega/2$.
(d) Frequency difference (magnitude) between the peak position and the trap laser frequency. The slower the lattice moves, the more biased is the momentum distribution of the nonlocalized atoms.
(e) The velocity distributions of the nonlocalized atoms are illustrated at various lattice speed. 
Black curve represent the atoms under momentum diffusion without being localized in the lattice potential minima.
Red solid(dashed) curve represents the nonlocalized(localized) atoms in the lattice at an intermediate lattice speed.
Blue solid(dashed) curve represents the nonlocalized(localized) atoms in the lattice at a slow lattice speed.
}
\label{fig2}
\end{figure*}

By fitting the observed spectrum with the appropriate profiles, the proportions of the localized atoms as well as the nonlocalized atoms in the lattice is obtained, respectively.
The number of the localized(nonlocalized) atoms in the optical lattice is proportional to the area under the inner three peaks(the outer two broad peaks).
It is observed that as the lattice speed increases the peaks at both ends gets larger [Fig.\ \ref{fig1}(b)], which means the fraction of the atoms nonlocalized in the lattice increases.
	
Note that the peak heights of the nonlocalized atoms are different. This can be understood in the following way.
As shown in Eq.~(\ref{dz_MOT}) in Methods, the center of atomic cloud is shifted to a nonzero magnetic field region, where atoms experience the Zeeman shifts. 
As a result, the effective detunings of the ${\sigma}^{+}$ and ${\sigma}^{-}$ trap beams experienced by the atoms in various ground-state magnetic sublevels are different, making the scattering rates of the ${\sigma}^{+}$ and ${\sigma}^{-}$ polarization lasers differ significantly.
Moreover, the scattered light by the ${\sigma}^{+}$ and ${\sigma}^{-}$ trap beams are in different elliptical polarizations while the detectors are measuring only the polarization in a particular direction (horizontal polarization in the lab coordinates), further deepening the discrepancy.
A detailed analysis on the asymmetric peak heights is given in Ref.\cite{jrjr}.
	
{\bf Signature of the localized atoms in the moving lattice.  } 
In our experiment, the optical lattice moves at the speed of $v_l=\Delta \omega/2k$ in the direction of the trap laser beam with the higher frequency $\omega+\Delta \omega$, where $\omega$ is the frequency of the lower-frequency trap laser beam and $k=\omega/c$ with $c$ the speed of light.
If the atoms move along with the lattice while localized at the lattice potential minima, the higher(lower) frequency trap laser is Doppler shifted by $-kv_l=-\Delta\omega/2$($+kv_l=+\Delta\omega/2$) in the rest frame of the moving atoms. In the Cartesian coordinates depicted in Fig.\ \ref{fig6}(a) in Methods, the higher frequency trap laser beams propagate in $x$,$y$ and $z$ directions, respectively, and therefore the resulting lattice velocity points to (1,1,1) direction, which is orthogonal to (-1,1,0) direction leading to our detector. Therefore, the light resonantly scattered from all six trap beams by these atoms into the detector direction would have the same frequency of $\omega+\Delta\omega/2$.
The Rayleigh peaks in the spectra shown in Fig.\ \ref{fig2}(a), occurring at this frequency, thus represent the atoms moving along with the lattice. 
The Rayleigh peak and two Raman sidebands on both sides are the signature of the atoms in LDR, confined in a space whose dimension is less than the wavelength $\lambda$, and therefore together they represent the atoms localized in the potential minima of the moving optical lattice.
\begin{figure*}
\centering
\includegraphics[width=\textwidth]{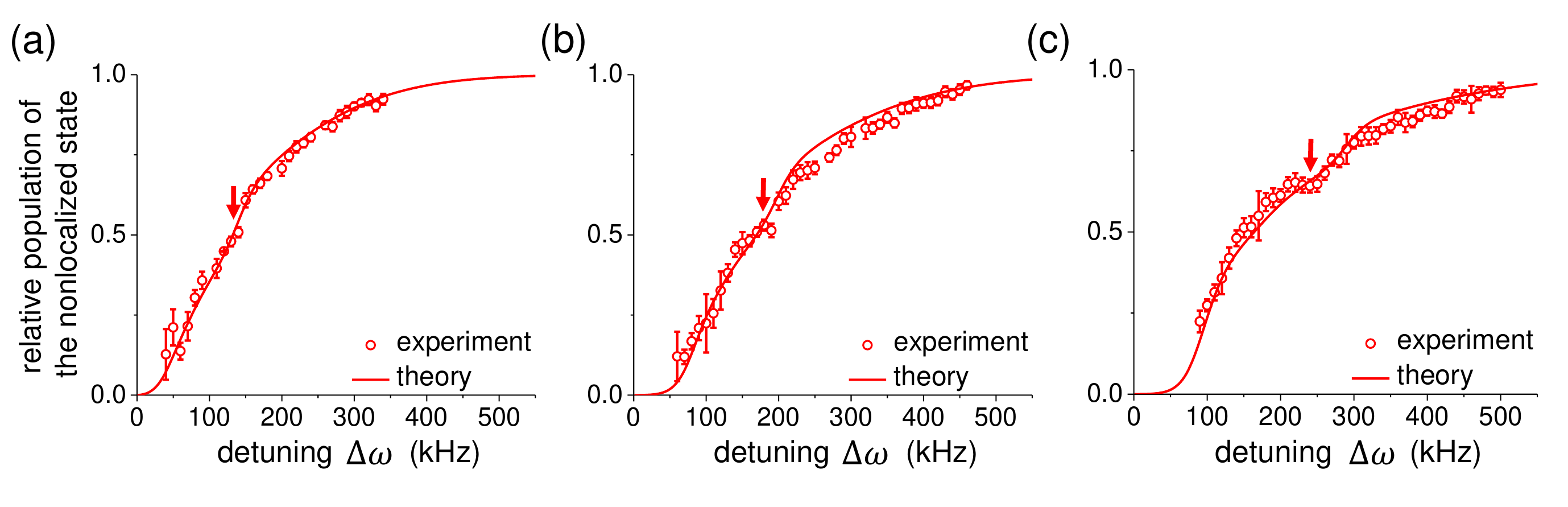}
\caption{{\bf Relative population of the nonlocalized atoms at various trap laser intensity.} Red dots represent the populations from the experimental observation at various potential height $U_0$ and the trap laser intensity $I_0$. (a) $U_{0}=840E_{r}$ with $I_{0}=$0.89mW/cm$^{2}$, (b) $U_{0}=1600E_{r}$ with $I_{0}=$1.8mW/cm$^{2}$, and (c)  $U_{0}=2400E_{r}$ with $I_{0}=$2.7mW/cm$^{2}$, where $E_{r}$ is the photon-recoil energy with a value of $E_{r}/\hbar=3.86$kHz. Error bars are derived from the fitting errors as in Fig.~\ref{fig1}(a). Stepwise increase in population is observed in experiment. 
Red solid curves are the solutions of the rate equation, Eqs.~(1)-(3).
The stepwise increase is shown to be due to the characteristic transmittance of the first-excited vibration state localized at an potential minimum of the moving lattice.
}
\label{fig3}
\end{figure*}
		
{\bf Biased momentum distribution of the nonlocalized atoms.  } 
The atoms nonlocalized in the lattice still undergo continuous scattering of the trap laser beams, resulting in cooling of the atoms. In the steady state, therefore, these atoms acquire a certain momentum distribution\cite{Castin:89}.
Among these atoms, the atoms with their velocities close to the velocity of the lattice tend to be captured in the lattice and thus the distribution near the lattice velocity is decreased. 
As a result, the velocity distribution of the nonlocalized atoms is biased in the opposite direction to the traveling direction of the lattice.
Figure \ref{fig2}(c) shows the experimental evidence of this bias. 
The broad peaks at both sides in the spectrum represents the resonance fluorescence of the nonlocalized atoms excited by the trap lasers with the frequencies $\omega$ and $\omega+\Delta \omega$($\Delta \omega /2\pi=280$kHz), respectively. 
The dashed red vertical line represents the frequency of trap beam at $\omega$. 
If there is no bias in the atomic velocity distribution, the pick should be centered around this line. 
However, the peak is centered around the dashed blue line biased by about 47kHz to the red. 
In addition, the peak on the right side is biased by about 51kHz to the blue. 
Both observations indicate that the velocity distribution of the nonlocalized atoms is biased in the opposite direction to the lattice propagation direction. 

As seen in Fig.\ \ref{fig2}(d), the momentum bias is larger when the detuning of trap beam $\Delta \omega$, proportional to the speed of lattice, is smaller. 
We can explain this feature as follows. 
Suppose the optical lattice moves so fast that most of atoms are nonlocalized in the lattice while remaining as a cold gas in the background MOT. 
Their momentum distribution resulting from laser cooling would then be centered around $v=0$, corresponding to a black curve in Fig.\ \ref{fig2}(e). 	
Now consider the case where the lattice moves at an intermediate speed corresponding to the red arrow in Fig.\ \ref{fig2}(e).
The atoms with a velocity similar to the lattice speed would then be localized in the lattice (red dashed curve). 
As a result, the distribution around that velocity is reduced(red solid curve), inducing the distribution to be biased to the opposite direction.  
If the speed of the lattice is further reduce close to zero, most of the atoms would be localized in the lattice (blue dashed curve), leaving only the atoms moving fast in the opposite direction to the lattice propagation direction. As a result, the distribution of the nonlocalized atoms is strongly biased to the opposite direction (blue solid curve).

{\bf Evolution of the nonlocalized state population with the lattice speed.  }
The number of atoms localized in the lattice is proportional to the area under the Rayleigh peak and the Raman sidebands. Similarly, the number of nonlocalized atoms in the lattice is obtained from the area under the two Doppler-broadened peaks. The red dots in Figs.\ \ref{fig3}(a)-(c) represent the relative population of the nonlocalized atoms as a function of the lattice speed. 
The relative population of the nonlocalized atoms increases with the lattice speed. Interestingly, the population increases with a small stepwise jump at a particular lattice speed indicated by arrows. This tendency can be observed more clearly when the trap depth is larger (with larger trap laser intensity).
This interesting feature is due to the transmittance of the first excited vibration state of the atoms localized in one lattice node to neighboring ones as to be shown below. \\

\noindent
{\bf \large Discussion} 

{\bf A deterministic rate equation model of the atomic dynamics.  }
\label{sec-rateeq}
Theoretical methods such as the quantum Bloch equation\cite{Stenholm:85} and the Monte-Carlo wave function method\cite{Molmer:93} are usually employed in order to explain matter-wave effects in a modulated optical lattice. 
However, due to the inherent complexity of MOT, it is difficult to apply those methods directly to the present problem. Moreover, even if we were able to do so, it would be difficult to obtain a clear physical picture.  
Instead, we take a different approach. First, we expect the present system to be describable in terms of a non-equilibrium steady state with continuous transitions among the localized vibrational states of the lattice and the nonlocalized state. 
We recall that the deterministic rate equation models -- populations among states changing at deterministic rates -- are applied when describing non-interacting driven dissipative bosonic systems\cite{10.1038/ncomms7977} and molecular motors\cite{GE201287} in NESS's from a macroscopic point of view.
Similarly, we adapt a rate equation model to explain the experimental results in Figs. \ref{fig3}(a)-(c). 
	
\begin{figure*}
\includegraphics[width=0.65\textwidth]{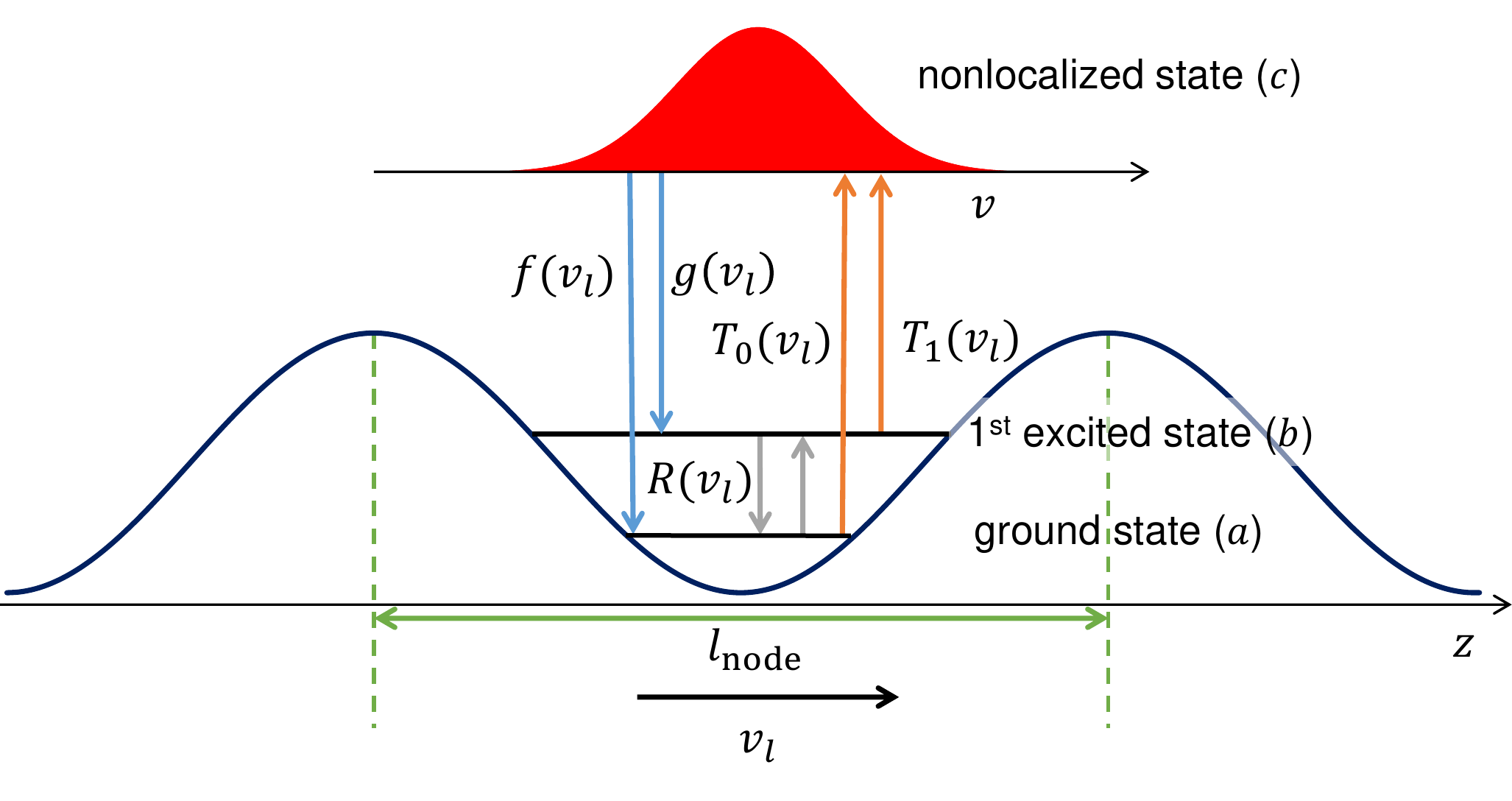}
\caption{{\bf Schematics of the rate equation model.} The lattice moves at a speed of $v_l$. Because of the periodicity of the lattice potential, we only consider one lattice period denoted by $l_{\rm node}$. By the same reason, we consider time evolution during $T=l_{\rm node}/v_l$. The populations in the ground(the first excited) vibration state is denoted by $a$($b$) while the population in the nonlocalized state is expressed as $c$.
The rates $f(v_l)$ and $g(v_l)$ and the transition amplitudes $T_0(v_l), T_1(v_l)$ and  $R(v_l)$ are defined in the text.
}
\label{fig4}
\end{figure*}
	
Our rate equation model describes the atomic transitions among the localized and nonlocalized motional states of atoms (Fig.\ \ref{fig4}).
First, only the two lowest bound states of a lattice node is considered in the model 
because the band structure of the three-dimensional optical lattice generated by the present passively-stabilized MOT shows that only the ground and the first excited vibrational states are tightly bound\cite{wrkim2011}. 
The atoms in the localized states are strongly confined in the lattice potential minima. 
On the other hand, the nonlocalized atoms are considered to be in a cold-gas state trapped in the background MOT.
We consider the atomic motion in an one-dimensional lattice in the traveling direction of the optical lattice and neglect the atomic motion perpendicular the lattice motion. It is because the motion to other directions in the rest frame of the moving optical lattice is suppressed by high potential barriers.
In addition, the interstate transition rates in our rate equation are pre-averaged over a time period $T=l_{\rm node}/v_{l}$, which is the time taken for the lattice to move one lattice period $l_{\rm node}$. 
The reason for this averaging is 
because the lattice potential is periodic and thus the rapid atomic dynamics during the time period $T$ is continuously repeated as the lattice moves.
By such pre-averaging, we can eliminate rapid changes occurring in the time scale of $T$ in the rate equation and can deal with the slow evolution of the state populations.  By the same reason, we only consider one lattice period denoted by $l_{\rm node}$ as shown in Fig.~\ref{fig4}.
The rate equation can then be written as
\begin{eqnarray}
	\dot{a} &=& -\frac{1}{T}\left[R(v_l)+T_{0}(v_l)\right]a+\frac{R(v_l)}{T} b+g(v_l)c\\
	\dot{b} &=& \frac{R(v_l)}{T} a-\frac{1}{T}\left[T_{1}(v_l)+R(v_l)\right]b+f(v_l)c\\ 
	\dot{c} &=& \frac{1}{T}\left[T_{0}(v_l)  a+T_{1}(v_l) b\right]-\left[f(v_l)+g(v_l)\right]c
\end{eqnarray}
where $a$ and $b$ are the populations of the ground and the first excited bound states of the lattice potential as shown in Fig.\ \ref{fig4}, respectively, and $c$ is the population in the nonlocalized state. 
We assume that nonlocalized atoms have a certain momentum distribution corresponding to the observed Doppler-broadened peaks in the spectrum. The quantities $T_{0}(v_l)$ and $T_{1}(v_l)$ are the transition amplitudes from the ground(the first excited) vibrational state to the nonlocalized state  during the time period $T$.
The rates $f(v_l)$ and $g(v_l)$ are the transition rates from the nonlocalized state to the ground and the first excited vibrational states, respectively, whereas $R(v_l)$ is the transition amplitude between the two vibrational states during the time period $T$. 
Explicit formulae for these parameters are derived in Methods.
Because what we observed in the experiment corresponds to a steady state, we solve the rate equation for the steady state by letting $\dot{a}=\dot{b}=\dot{c}=0$.

\begin{figure*}
\includegraphics[width=0.97\textwidth]{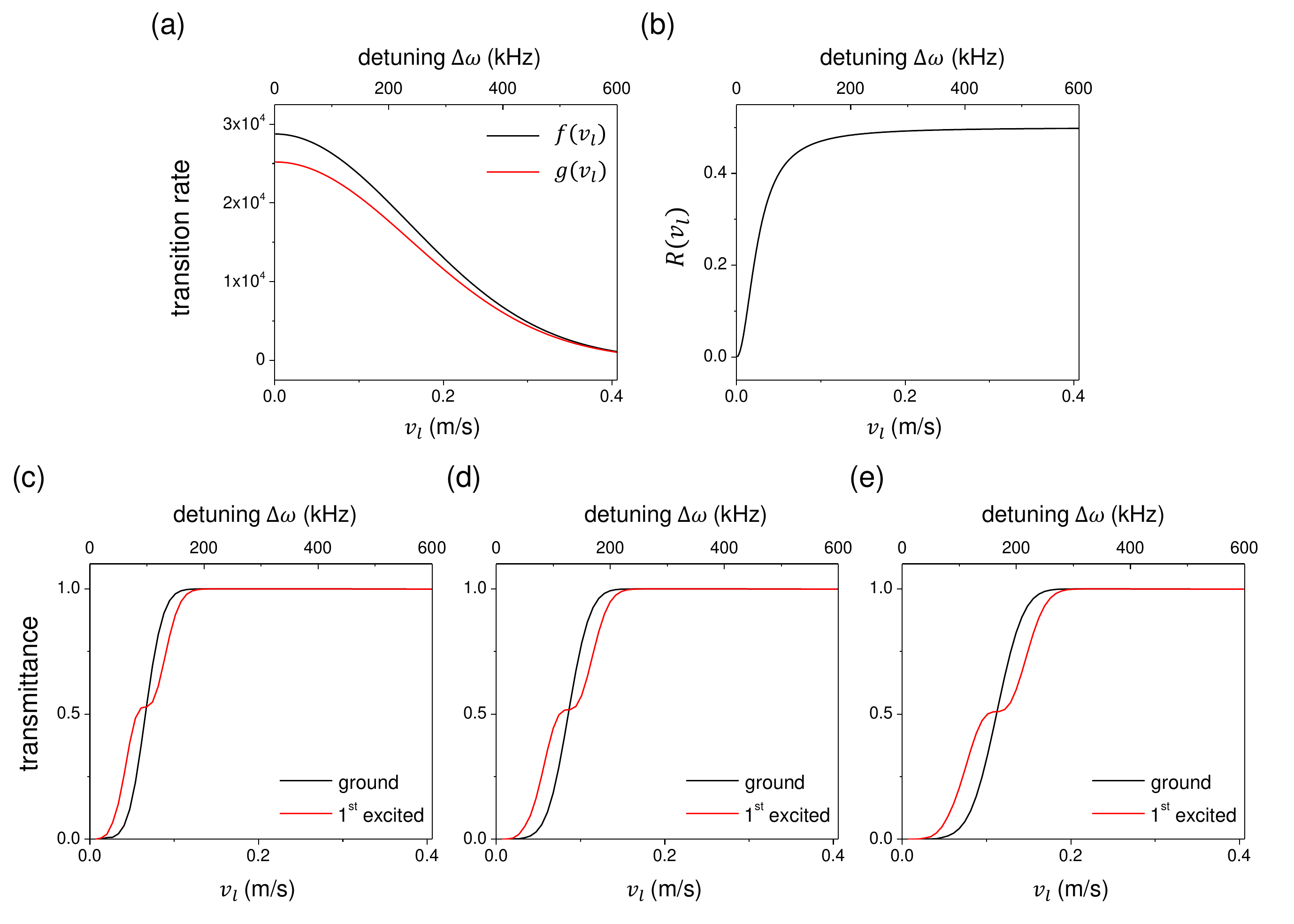}
\caption{{\bf Variation of the calculated transition rates and transition amplitudes with the lattice speed.}
(a) Transition rates $f(v_{l})$ and $g(v_{l})$ calculated by using Eqs.\ (\ref{fvl}) and (\ref{gvl}) in Methods, respectively. (b) Transition amplitude $R(v_{l})$ between two bound states of the lattice during the time period $T$, calculated by using Eq.\ (\ref{Rvl}). 		
Lattice potential height $U_{0}=1600E_{r}$ corresponding to $I_{0}=$1.8mW/cm$^{2}$ and the oscillation frequency $\omega_{\rm osc}/2\pi$ =91kHz are used with a free parameter $\eta$=$3\times10^7$/s for numerical calculations.
(c)-(e) Transition amplitude or transmittance of the atoms in the bound states during the time period $T$ as the lattice speed increases in the case of three different potential heights. Parameters used in the calculations are as follows: (c) $U_{0}=840E_{r}$, (d) $U_{0}=1600E_{r}$, and (e) $U_{0}=2400E_{r}$. The black curve represents the transmittance of the ground state and the red curve corresponds to the first excited state. Transmittance shape gets extended as the depth of the optical lattice potential increases.
}
\label{fig5}
\end{figure*}

\begin{figure*}		
\includegraphics[width=\textwidth]{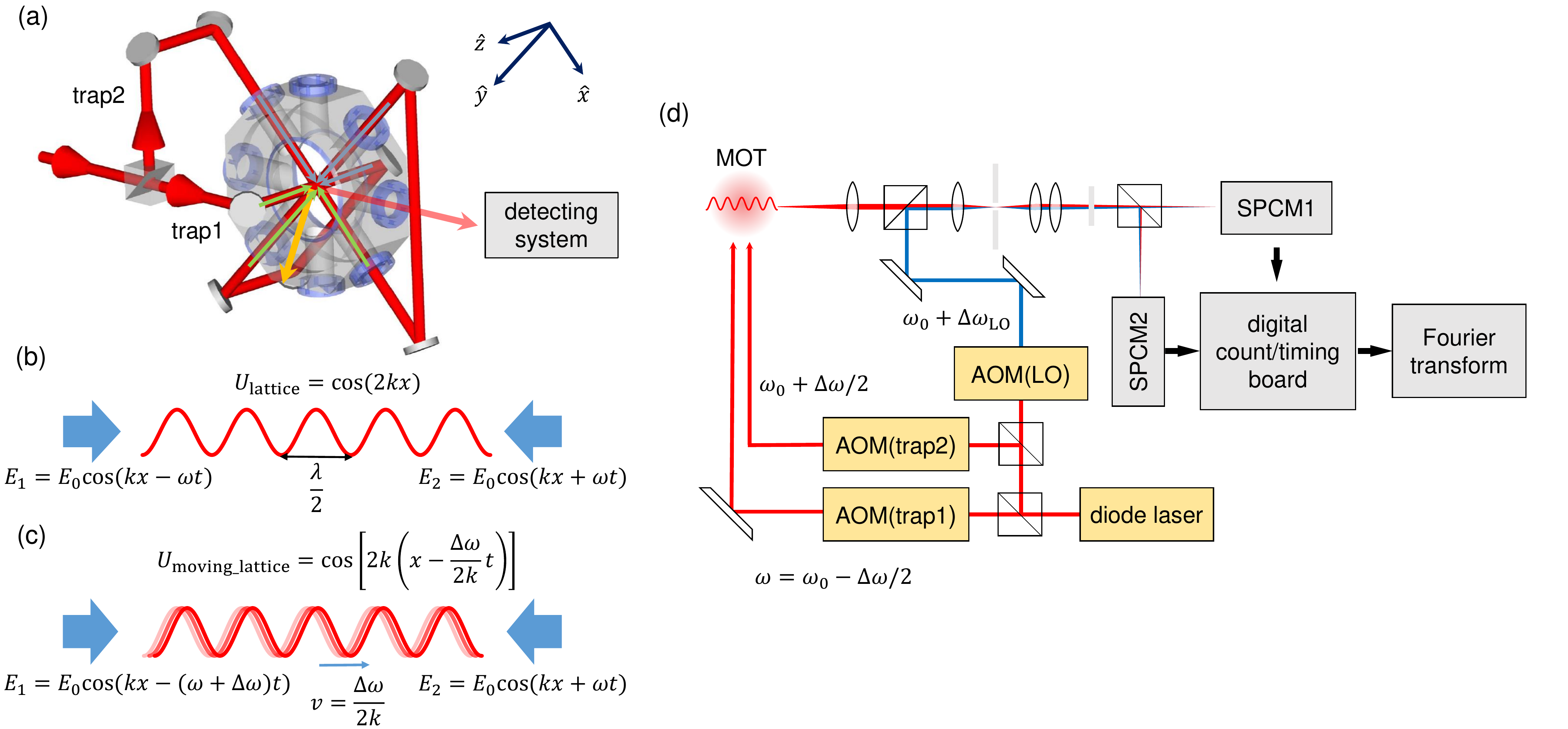}
\caption{{\bf Passively stabilized MOT with a moving optical lattice and a setup for measuring resonance-fluorescence spectrum.}
(a) Three-dimensional laser configuration for a moving optical lattice in a phase-stabilized MOT. The blue arrows represent the trap laser with frequency of $\omega+\Delta \omega$ while the green arrows indicate the trap laser with frequency of $\omega$. The yellow arrow represents the direction of the moving optical lattice. In the coordinate system shown, it is in (1,1,1) direction. Fluorescence is collected in (-1,1,0) direction.
(b) Configuration of laser beams for an one-dimensional static optical lattice. 
(c) The same for an one-dimensional moving optical lattice. 
(d) Experiment schematic for measuring resonance fluorescence spectrum of atoms. A hybrid of a moving optical lattice and a background MOT is generated in a phase-stabilized MOT with two bichromatic lasers whose frequencies are detuned by AOM's. 
Spectrum measurements are done by using PCSOCS on the fluorescence mixed with a local oscillator(LO). 
The diode laser frequency $\omega_{0}=\omega+\Delta\omega/2$ is detuned by -3$\Gamma$ from the atomic resonance. 
The detuning for the local oscillator is $\Delta \omega_{\rm LO}$=-10MHz.
}
\label{fig6}	
\end{figure*}

	
As explained in the previous section, the nonlocalized atoms have a biased momentum distribution due to the sub-Doppler cooling by the moving lattice. 	
The atoms which have velocities around the moving lattice speed are cooled by the lattice and then become localized in the lattice. The probability of the atoms being cooled to the bound states of the lattice would decrease as the lattice speed increases. This trend is reflected in the transition rates $f(v_l)$ and $g(v_l)$ as shown in Fig.\ \ref{fig5}(a). They were calculated by using Eqs.\ (\ref{fvl}) and (\ref{gvl}) in Methods, respectively.
	
The atoms in the bound states of the lattice would undergo mutual transitions between the bound states when the lattice is translated rapidly. We expect the transition amplitude $R(v_l)$ would increase as the lattice speed goes up and eventually reach 0.5 at very high lattice speeds. This is exactly what we observe in the numerically calculated $R(v_l)$ as shown in Fig.\ \ref{fig5}(b).
	
	
When the lattice moves, some atoms localized in the lattice would tunnel through or go over the potential barrier to the neighboring lattice nodes.
These atoms are no longer localized in the potential minima of the lattice and thus considered to be `nonlocalized'.
They will end up in the cold gas in the background MOT.
This transition to the nonlocalized state is closely related to the momentum distribution in the bound states. 
If the momentum distribution has a large component in the opposite direction to the lattice motion, the transition amplitude would be high.
The probability amplitude of the ground state has a Gaussian distribution and thus its momentum distribution is also a Gaussian centered at zero momentum. 
The ground state is thus expected to show a smooth transition to the nonlocalized state as the lattice speed increases.

\begin{figure*}
\centering
\includegraphics[width=0.8\textwidth]{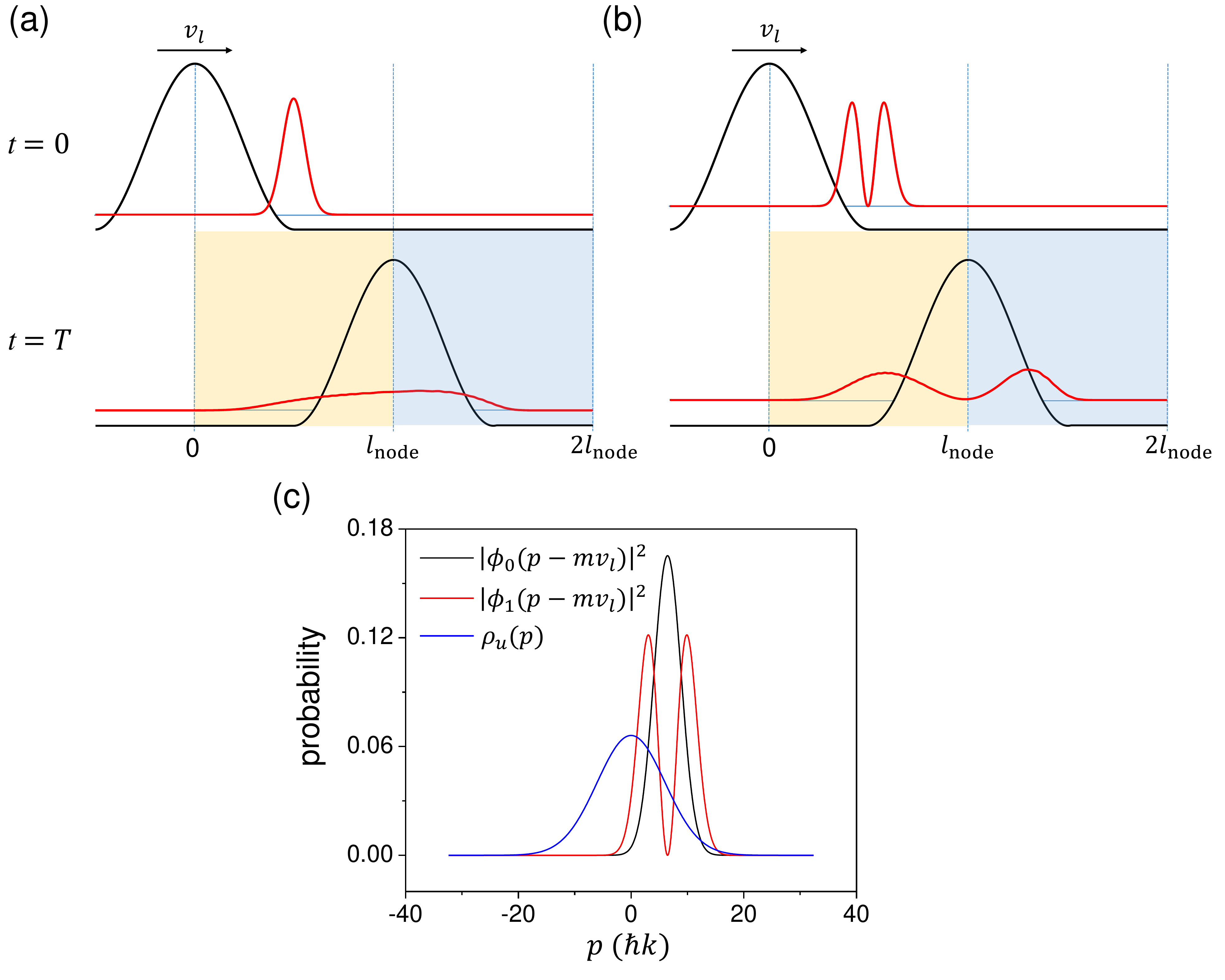}
\caption{{\bf Transmittance calculation example and momentum distributions of localized as well as unlocalized atoms.}
(a) Calculating transmitted waves (red curves) through a moving potential barrier (black curve). The initial wave function is the ground state and (b) it is the first excited state of a harmonic oscillator.
The probability amplitude in the region between $x=0$ and $l_{\rm node}$ is considered being transmitted. 
In this example, the following parameters were used: 
$V_0=1600E_{r}$, $v_l=0.086$m/s and $\omega_{\rm osc}/2\pi=91$kHz.
(c) The momentum distribution $\rho_{u}(p)$ of the unlocalized atoms under momentum diffusion, compared with the momentum distributions $|\phi_{0,1}(p-mv_{l})|^{2}$ of the localized atoms in the ground and the first excited states of a local potential minimum of the optical lattice. 
The distribution $\rho_{u}(p)$ is estimated from the spectrum at a large trap laser detuning of $\Delta \omega/2\pi$=500kHz, with most of the atoms in the background MOT. The distribution corresponds to the atomic temperature of about 35$\mu$K.
In this example, the lattice moves with a speed of $v_{l}$=0.067m/s with the trap laser detuning of $\Delta \omega/2\pi$=100kHz.
}
\label{fig7}
\end{figure*}

On the other hand, the probability amplitude of the first excited state has a node in the center and its momentum distribution has two opposite momentum components as shown in Fig.~\ref{fig7}(c) in Methods. As the lattice speed increases, the momentum component opposite to the lattice motion would undergo the transition to the nonlocalized state first. 
The momentum component in the same direction as the lattice motion would do so only when the lattice speed is larger enough. 
As a result,  for the first excited vibrational state we expect to see a stepwise transition to the nonlocalized state as the lattice speed increases.
This trend is confirmed in the numerically calculated transition amplitudes $T_0(v_l)$ and $T_1(v_l)$ as shown in Figs.\ \ref{fig5}(c)-(e).
		
	
The stepwise transmittance of the first excited state is strongly reflected in the evolution of the nonlocalized state population as the lattice speed increases. Our rate equation model well reproduces the observed stepwise evolution of the nonlocalized state population as shown in Fig.\ \ref{fig3} at three different potential heights. This agreement supports the validity of our rate equation model.

{\bf Evidences for an NESS.  }
Based on the above analysis of the atomic dynamics, we conclude that our system is in an NESS. 
When a system is in an NESS, not only the system is in a steady state but also there must be non-vanishing currents among possible states.
In our case, both conditions are satisfied.
First, we know that the atoms are in a steady state because we can observe a stationary atom cloud as well as the spectrum of resonance fluorescence of atoms. 
We also know that the total number of atoms in the hybrid trap composed of the optical lattice and the background MOT is nearly unchanged when we vary $\Delta\omega$ from zero to 500kHz as in Figs.~\ref{fig2}, \ref{fig3} and \ref{fig5}.
It is because the trap laser frequency is not changed much compared to the overall red detuning of 3$\Gamma/2\pi$=18MHz. 
The difference in the scattering rates of these frequency-detuned trap lasers is at most 0.9\%.

Next, the biased momentum distribution of atoms in the nonlocalized state is a strong evidence that the unlocalized 
atoms are recaptured by stochastic momentum diffusion to the localized state in the optical lattice nodes. 
In other words, we have a continuous current of atoms from the unlocalized to the localized states.
We also have a current in the opposite direction. 
Otherwise, 
all of the atoms localized in the moving lattice would move to the edge of the MOT and then escape from the MOT. 
This would result in a significant reduction of the total number of atoms in the hybrid trap when we move the optical lattice.
Instead, we observed the total number of atoms almost unchanged.
This indicates that the localized atoms in the lattice continuously undergo a transition to the nonlocalized state.
The relative strength of these currents in opposite directions determine the relative populations in the localized and the unlocalized states as a function of the lattice speed.
From these considerations, we conclude that our system is in an NESS\cite{1742-5468-2007-07-P07012}.
	
{\bf Connection to ASEP.  }
Generally, an NESS system is associated with a stochastic process for particles. 
In some cases, an additional external drive is present\cite{ZHANG20121}. 
In our experiment, atomic momentum change due to light scattering of trap lasers can be considered as a stochastic process, and the moving lattice potential can be regarded as an external drive. 	
Particularly, our system approximately suits the conditions of ASEP\cite{1742-5468-2007-07-P07023}.
In ASEP, the particles in a lattice hop from one node to neighboring nodes asymmetrically, biased by an external drive. Moreover, hopping to a neighboring node already occupied by a particle is suppressed.
In our case, tunneling or hopping rates to neighboring nodes are asymmetric because of the lattice motion. 
Moreover, only one atom can occupy a local minimum of the lattice potential because of the high light-assisted two-atom collision rate ($\sim 10^8$/s at least)\cite{PhysRevA.76.013402} in each small volume of the local minima of the lattice potential.

When ASEP is combined with the Langmuir kinetics\cite{PhysRevE.70.046101}, it can describe particle absorption from and emission to a reservoir of absorbates.
In our system, 
the moving optical lattice plays a role of an external drive as well as absorbing sites and the nonlocalized atoms in the background MOT serve as absorbates.
Under this setting, 
some atoms are localized at the potential minima of the optical lattice while some tunnel through the potential barrier to the adjacent potential minima. Moreover,  some atoms in the nonlocalized state are captured to the optical lattice (absorption of absorbates), resulting in a biased momentum distribution of the atoms in the nonlocalized state. We also have some of the localized atoms undergo a transition to the nonlocalized state (emission to the reservoir). 
These considerations indicate that our system is in an NESS with characteristics of ASEP with Langmuir kinetics.
Therefore, our system can be used to simulate the NESS's in various biochemical systems\cite{PhysRevLett.93.238102,PhysRevE.95.012113}.
Interestingly, our system supports quantum internal states of the lattice nodes (vibrational states), which lacks in the standard ASEP model, and thus our system can be regarded as an extended model suitable for low temperature.
More detailed analysis of our systems in the viewport of ASEP is beyond the scope of the present study but would be interesting to perform in the future.\\

\noindent
{\bf \large Methods}

{\bf Generation of a moving optical lattice in a hybrid trap.  } 
\label{gmlo}
In a phase-stabilized MOT, two phase-related trap beams propagating in the opposite directions are folded to form three sets of counter-propagating laser beams as shown in Fig.\ \ref{fig6}(a). 
In the case of a static optical lattice, the trap lasers have the same frequency detuning of -3$\Gamma$ from the (F=3) $\leftrightarrow$ (F=4) transition of $^{85}$Rb D2 line, where $\Gamma/2\pi$=6MHz is the decay rate of the transition.
The interference pattern forming the optical lattice is stable due to the time-phase relation of two trap beams\cite{wrkim2011}. 
In our experiments, the trap depth of the lattice potential can be adjusted by the laser intensity between 840$E_{r}$ to 2400$E_{r}$, where $E_{r}$ is the recoil energy with a value of $E_{r}/\hbar=3.86$kHz. The atoms then experience the sub-Doppler cooling due to the polarization gradient of the trap beams\cite{Dalibard:89}
and become localized at the potential minima of the optical lattice.	
In most of our experiments, the trap depth was 1600 $E_{r}$ and the oscillation frequency $\omega/2\pi$ of the generated lattice potential was 91kHz. 	
	
We can move the optical lattice by changing the frequency of one of the trap laser beams. 
With a frequency difference $\Delta\omega$ between them, 
the optical lattice moves in the direction of the higher frequency trap laser
with a speed of $v_{l}=\Delta\omega/2k$, where $k=2\pi/\lambda$ is the wave number and $\lambda$ is the trap laser wavelength [see Fig.\ \ref{fig6}(c)]. 
We used acousto-optic modulators(AOM's) to shift the trap laser frequencies. In the three-dimensional case of Fig.\ \ref{fig6}(a), the optical lattice moves with an equal velocity component of $v_{l}$ along each axis, resulting in the lattice velocity in (1,1,1) direction in the Cartesian coordinate system.
	
The MOT force experienced by atoms in this trap is different from that of the conventional MOT. 
The force in the conventional one-dimensional model of MOT with the same trap laser detunings can be described as\cite{atomics},
\begin{align}
F_{\rm MOT} = -\alpha v-\frac{\alpha\beta}{k} z,
\label{c_MOT}
\end{align}  
where $v$ and $z$ are the velocity and position of an atom, respectively, and $\alpha$ and $\beta$ are the constants determined by the MOT parameters.
If the frequencies of the trap beams are different by $\Delta \omega$, the force of MOT changes to
\begin{align}
F'_{\rm MOT}=-\alpha v-\frac{\alpha\beta}{k} (z-z_{0}),
\label{dz_MOT}
\end{align}
where 
$z_0\equiv \Delta\omega/2\beta$. 
Due to the modified restoring force given by the second term in Eq.\ (\ref{dz_MOT}), the near-Doppler-temperature atoms would then be trapped at a shifted trap center $z_{0}$ while the sub-Doppler-cooled low-temperature atoms would be localized at the potential minima of the moving optical lattice. 
	
{\bf Spectrum measurement. } 
\label{sm}
In our experiments, the MOT with the moving optical lattice was loaded and then the fluorescence spectrum of trapped atoms exited by the trap laser was measured. We used a ultrasensitive heterodyne spectroscopic technique called the photon-counting-based second-order correlation spectroscopy(PCSOCS)\cite{Hong:06}. 
In this technique, the atomic fluorescence is mixed with a local oscillator laser and 
all of the arrival times of the photons of the combined light are recorded on two single-photon counting modules (SPCMs) as shown in Fig.\ \ref{fig6}(d).

The photon arrival time record is then used to obtain the second-order correlation function $g^{(2)}(\tau)$, and the spectrum can then be obtained by Fourier transforming it. 
Typical photon count rate registered on each detector was about $10^6$cps and the fluorescence was measured for about an hour for enough signal-to-noise ratios. 
In order to make both the magnitude of the lattice potential and the lattice velocity well defined, it was essential to stabilize the frequency as well as the intensity of the trap laser.
We employed a proportional-integral-derivative(PID) feedback control in stabilizing the intensity and frequency of the trap laser. The intensity of the laser was stabilized within 5\% of its maximum, and the frequency of the laser was stabilized within 5kHz, which mainly contributed to the resolution of the spectrum.

{\bf Formulae for the parameters used in the rate equation model.  }
 \label{rate}
Transition amplitudes
$T_{0}(v_l)$ and $T_{1}(v_l)$ from the ground and the excited vibrational states, respectively, to the nonlocalized state are obtained 
by calculating the transmittance of the corresponding vibrational state through the potential barrier as the lattice potential moves during time $T=l_{\rm node}/v_l$. 
We solve the time-dependent Schr\"{o}dinger equation for time $T$ with an initial wave function $\psi_{0,1}(x,t=0)$ placed at a minimum of the lattice potential. 
\begin{align}
	i\hbar\dfrac{\partial\psi}{\partial t}=-\frac{\hbar^{2}}{2m}\frac{\partial^2\psi}{\partial x^{2}}+V(x-v_{l}t), \qquad (0<t<T) 
\end{align}
where $V$ describes an one-dimensional lattice potential moving at a speed of $v_{l}$ with a lattice period of $l_{\rm node}$ = $\sqrt{3}/2 \lambda$ and $m$ is the atomic mass. 
For the initial wave functions, we use the ground and the first excited states of a simple harmonic oscillator with the same oscillation frequency as our experiment.
\begin{table*}
\centering
\begin{tabular}{| c | c | c | c | c | c | c| c|}
\hline
Trap power & Trap depth &$\omega_{G}$&$\omega_{L}$&$\omega_{\rm osc}/2\pi$& $v_{le}$ & $\Delta \omega_{le}/2\pi$ & $\eta$ \\ \hline
 0.89mW/cm$^{2}$& 840$E_{r}$ &119kHz&77kHz&67kHz& 0.069m/s & 103kHz &$2.5\times10^{7}$/s \\ \hline
 1.8mW/cm$^{2}$& 1600$E_{r}$ &89kHz&110kHz&91kHz& 0.096m/s & 142kHz &$3.0\times10^{7}$/s \\ \hline
 2.7mW/cm$^{2}$& 2400$E_{r}$ &131kHz&161kHz&99kHz& 0.12m/s & 174kHz &$3.1\times10^{7}$/s \\ \hline
\end{tabular}
\caption{Several experimental parameters used to determine the values of  $f(v_l)$, $g(v_l)$, $T_{0}(v_l)$, $T_{1}(v_l)$ and $R(v_l)$ in the rate equation, Eqs.~(1)-(3).
Parameters $\omega_{G}$ and $\omega_{L}$ are the Gaussian and Lorentzian widths of a Voigt profile, respectively, fitting the observed spectrum at a high lattice speed ($\Delta \omega/2\pi$=500kHz) with most of the atoms trapped in the background MOT. 
The oscillation frequency of the potential minima of the optical lattice is denoted by $\omega_{\rm osc}$.
The parameter $v_{le}$ is the lattice speed at which an atom initially stationary can go over the potential barrier classically. It is given by $v_{le}=\sqrt{2V_0/m}$ with $V_0$ the potential depth. The trap laser detuning corresponding to $v_{le}$ is denoted by $\Delta \omega_{le}$. The parameter $\eta$ is a fitting parameter used in calculating $f(v_l)$ and $g(v_l)$.
}
\label{forT}
\end{table*}

Since we are interested in the change of the wave function in one lattice period, we approximate the periodic potential with a moving barrier corresponding to one period of the optical lattice potential as shown in Figs.~\ref{fig7}(a) and (b). At $t=0$, the center of the potential barrier is located at $x=0$ and the wave function is centered around $x=l_{\rm node}/2$. Transmittance is obtained by integrating the absolute square of the wave function between $x=0$ and $x=l_{\rm node}$ at time $t=T$.
\begin{align}
\label{T0T1}
T_{0,1}(v_l) = \int_{0}^{l_{node}}|\psi_{0,1}(x,T)|^{2}dx
\end{align}

\noindent
Calculation of transmittance was done numerically\cite{doi:10.1119/1.4833557}. We set the `transmitted' atoms as the atoms to be found in the neighboring potential minimum corresponding to a region shaded in yellow in Figs. \ref{fig7}(a) and (b).
Because we are interested in the time evolution during $T$ with a periodic potential moving by one lattice period, 
we consider only the transmitted wave in that region.
Examples of calculated $T_{0}(v_l)$ and $T_{1}(v_l)$ are shown in Figs.~\ref{fig5}(a)-(c) as a function of $\Delta \omega$ (or $v_l=\Delta \omega/2k$).

The rates
$f(v_l)$ and $g(v_l)$ shown in the Fig.~\ref{fig5}(a) are the transition rates from the nonlocalized state to the localized states, {\em i.e.}, the ground and the first excited states of the potential minimum, respectively.
The unlocalized atoms undergo momentum diffusion due to the continuous scattering of the trap laser light, resulting in 
a certain momentum distribution $\rho_{u}(p)$ affected by sub-Doppler cooling. 
The momentum distribution function $\rho_{u}(p)$ can be estimated from the spectrum at high lattice speeds with most of the atoms in the background MOT.
Momentum diffusion eventually drives atoms to be captured in the lattice. The transition rate to a localized state is proportional to the convolution of $\rho_{u}(p)$ and the momentum distribution $|\phi_{0,1}(p-mv_{l})|^{2}$ of the localized states  [Fig.\ \ref{fig7}(c)]. 
Thus, $f(v_l)$ and $g(v_l)$ can be expressed as
\begin{align}
\label{fvl}
f(v_{l})=\eta\int_{-\infty}^{\infty}\rho_{u}(p)|\phi_{0}(p-mv_{l})|^{2}dp
\end{align}
\begin{align}
\label{gvl}
g(v_{l})=\eta\int_{-\infty}^{\infty}\rho_{u}(p)|\phi_{1}(p-mv_{l})|^{2}dp
\end{align}
where $\eta$ is a constant related to the momentum diffusion. The rates $f(v_{l})$ and $g(v_{l})$ decrease as the lattice speed increases, which means that the atoms tend to be nonlocalized as the lattice speed increases. The parameter $\eta$ takes a role of a fitting parameter in the rate equation model, and its value is listed in Table \ref{forT}.
		
The transition amplitude $R(v_l)$ between the ground and the first excited vibrational states during the time period $T$ is obtained by approximating a node of the moving optical lattice with a harmonic oscillator with the same oscillation frequency moving at the same speed. 
The equation of motion for the probability amplitudes in the case of a moving harmonic oscillator of an oscillation frequency $\omega_{\rm osc}$ is given by\cite{QM}, 
\begin{align}
\dot{a_{n}} = a_{n+1} v \sqrt{\dfrac{(n+1)m\omega_{\rm osc}}{2\hbar}}e^{-i\omega_{\rm osc} t}-a_{n-1}v\sqrt{\dfrac{nm\omega_{\rm osc}}{2\hbar}}e^{i\omega_{\rm osc} t}
\end{align}
where
$a_{n}$ is the probability amplitude of the $n$th state.
We consider the transition between the first two vibrational states. 
The result is
\begin{align}
 |a_{1}(t)|^2= \dfrac{1}{1+\frac{\hbar\omega_{\rm osc}}{2mv^{2}}}\sin^{2}\left(\dfrac{t}{2}\sqrt{2m\omega_{\rm osc} v^{2}/\hbar+\omega_{\rm osc}^{2}}\right)
 \label{Rv}
\end{align}
for initial condition $a_0(0)=1$ and $a_1(0)=0$. The expression for $|a_{0}(t)|^2$ has the same form as $|a_{1}(t)|^2$ in Eq.~(\ref{Rv}) for initial condition $a_0(0)=0$ and $a_1(0)=1$, and thus a single parameter $R(v_l)$ is sufficient.
It is obtained by taking a time average of the transition probability,
\begin{align}
\label{Rvl}
R(v_l)= \frac{1}{T}\int_0^T |a_{1}(t)|^2 dt\simeq \dfrac{1}{2(1+\frac{\hbar\omega_{\rm osc}}{2mv_l^{2}})}.
\end{align}

Table \ref{forT} lists the values of
several parameters obtained from experimental results at various trap laser intensities. 
These parameter values are used to calculate all of the parameters appearing in the rate equation.
The momentum distribution $\rho_{u}(p)$ is specified by a Voigt profile with a Gaussian width $\omega_{G}$ and a Lorentzian width $\omega_{L}$. 
The eigenstates of the harmonic oscillator is determined by the oscillation frequency $\omega_{\rm osc}$. From the eigenstates, the momentum distribution
$\phi_{0,1}(p)$ can be obtained.
The transition rates $f(v_l)$ and $g(v_l)$ are then given by 
Eqs.~(\ref{fvl}) and (\ref{gvl})  
with $\eta$ to be determined by fitting the experimental results in Fig.~\ref{fig3}. 
The transition amplitude $R(v_l)$ is given by Eq.~(\ref{Rvl}) using $\omega_{\rm osc}$. 
For the calculation of the transmittance $T_{0,1}(v_l)$, the trap depth and the oscillation frequency values are used. 
The parameter $v_{le}$ in Table I is the lattice speed at which an atom initially stationary can go over the potential barrier classically. It is not needed to determine the parameters appearing in the rate equation, but its values are listed as a reference.  
It is noted that the transition from the localized to the unlocalized states occurs around $v_{le}$ or the corresponding detuning $\Delta \omega_{le}/2\pi$ in Figs.~\ref{fig3} and \ref{fig5}(c)-(e).

\hbox{}

\noindent\textbf{Acknowledgements}

\noindent
This work was supported by a grant from Samsung Science and Technology Foundation under Project No. SSTF-BA1502- 05 and by the Korea Research Foundation (Grant No.~2016R1D1A109918326).\\

\noindent\textbf{Author Contributions}

\noindent
K.O.C., J.-R.K., S.Y., S.K. and K.A. conceived the experiment. 
K.O.C., J.-R.K. S.K. and J.K. performed the experiment. K.O.C. and J.-R.K. analyzed the data and carried out theoretical investigations. K.A. supervised overall experimental and theoretical works. K.O.C. and K.A. wrote the manuscript. All authors participated in discussions. K.O.C. and J.-R.K. equally contributed to the work.\\

\noindent\textbf{Additional Information}

\noindent\textbf{Competing Financial Interests }  The authors declare no competing financial interests.\\


\newpage		

\begin{thebibliography}{40}
	\bibitem{Guidoni1999} Guidoni, L.\ \& Verkerk, P. Optical lattices: cold atoms ordered by light. J. Opt. B {\bf 1}, R23 (1999).
	
	\bibitem{Bloom2014} Bloom,\ B.\ J.\ \emph{et al}. An optical lattice clock with accuracy and stability at the $10^{-18}$ level. Nature (London) {\bf 506}, 71-75 (2014).
	
	\bibitem{PhysRevA.78.043413} Argonov,\ V.\ Y.\ \& Prants, S.\ V. Theory of dissipative chaotic atomic transport in an optical lattice. Phys.\ Rev.\ A.\ {\bf 78}, 043413 (2008).
	
	\bibitem{PhysRevE.89.012917} Horsley, E., Koppell, S.\ \& Reichl, L.\ E. Chaotic dynamics in a two-dimensional optical lattice. Phys.\ Rev.\ E.\ {\bf 89}, 012917 (2014).
	
	\bibitem{Hensinger2001} Hensinger, W.\ K.\ \emph{et al}. Dynamical tunnelling of ultracold atoms. Nature (London) {\bf 412}, 52-55 (2001).
	
	\bibitem{PhysRevLett.99.190405} Salger, T., Geckeler, C., Kling, S.\ \& Weitz,\ M. Atomic Landau-Zener Tunneling in Fourier-Synthesized Optical Lattices. Phys.\ Rev.\ Lett.\ {\bf 99}, 190405 (2007).
	
	\bibitem{Jordens2008} Jordens, R.,	Strohmaier, N., Gunter, K., Moritz, H.\ \& Esslinger,\ T. A Mott insulator of fermionic atoms in an optical lattice. Nature (London) {\bf 455}, 204-207 (2008).
	
	\bibitem{RevModPhys.86.779} Aoki, H.\ \emph{et al}. Nonequilibrium dynamical mean-field theory and its applications. Rev. Mod. Phys. {\bf 86}, 779-837 (2014).
	
	\bibitem{MOON20171}  Moon, G., Heo, M.-S., Kim, Y., Noh, H.-R.\ \& Jhe,\ W. Nonlinear, Nonequilibrium and Collective Dynamics in a Periodically Modulated Cold Atom System. Phys. Rep. {\bf 698}, 1-30 (2017).
	
	\bibitem{Bloch2012} Bloch, I., Dalibard, J.\ \& Nascimbene, S. Quantum simulations with ultracold quantum gases. Nat.\ Phys.\ {\bf 8}, 267-276 (2012).
	
	\bibitem{ZHANG20121} Zhang, X. -J., Qian, H.\ \& Qian, M. Stochastic Theory of Nonequilibrium Steady States and Its Applications: Part I. Phys. Rep. {\bf 510}, 1-86 (2012).
	
	\bibitem{Hsiang2015} Hsiang, J. -T.\ \& Hu, B.\ L. Quantum entanglement at high temperatures? Bosonic systems in nonequilibrium steady state. J. High Energy Phys. {\bf 2015}, 90 (2015).
	
	\bibitem{PhysRevLett.119.140602} Schnell, A., Vorberg, D., Ketzmerick, R.\ \& Eckardt,\ A. High-Temperature Nonequilibrium Bose Condensation Induced by a Hot Needle. Phys.\ Rev.\ Lett.\ {\bf 119}, 140602 (2017).
	
	\bibitem{PhysRevLett.93.238102} Kosztin, I.\ \& Schulten, K. Fluctuation-Driven Molecular Transport Through an Asymmetric Membrane Channel. Phys.\ Rev.\ Lett.\ {\bf 93}, 238102 (2004).
	
	\bibitem{PhysRevE.70.046101} Parmeggiani, A., Franosch, T.\ \& Frey, E. Totally asymmetric simple exclusion process with Langmuir kinetics. Phys.\ Rev.\ E.\ {\bf 70}, 046101 (2004).
	
	\bibitem{0034-4885-74-11-116601} Chou, T., Mallick, K.\ \& Zia, R.\ K.\ P. Non-equilibrium statistical mechanics: from a paradigmatic model to biological transport. Rep. Prog. Phys. {\bf 74}, 116601 (2011).
	
	\bibitem{PhysRevB.88.245114} Sabetta, T.\ \& Misguich, G. Nonequilibrium steady states in the quantum XXZ spin chain. Phys. Rev. B.\ {\bf 88}, 245114 (2013).
	
	\bibitem{PhysRevB.81.144301} Heyl, M.\ \& Kehrein, S. Nonequilibrium steady state in a periodically driven Kondo model. Phys. Rev. B {\bf 81}, 144301 (2010).
	
	\bibitem{PhysRevLett.111.240405} Vorberg, D., Wustmann, W., Ketzmerick, R.\ \& Eckardt, A. Generalized Bose-Einstein Condensation into Multiple States in Driven-Dissipative Systems. Phys.\ Rev.\ Lett.\ {\bf 111}, 240405 (2013).
	
	\bibitem{PhysRevLett.116.235302} Labouvie, R., Santra, B., Heun, S.\ \& Ott, H. Bistability in a Driven-Dissipative Superfluid. Phys. Rev. Lett. {\bf 116} 235302 (2016).
	
	\bibitem{Rauschenbeutel199845} Rauschenbeutel, A., Schadwinkel, H., Gomer, V.\ \& Meschede, D. Standing light fields for cold atoms with intrinsically stable and variable time phases. Opt. Commun. {\bf 148}, 45-48 (1998).
	
	\bibitem{Westbrook1997} Westbrook, C.\ I. \emph{et al}. A study of atom localization in an optical lattice by analysis of the scattered light. J. Mod. Opt. {\bf 44}, 1837-1851 (1997).
	
	\bibitem{PhysRev.89.472} Dicke, R.\ H. The Effect of Collisions upon the Doppler Width of Spectral Lines. Phys.\ Rev.\ {\bf 89}, 472-473 (1953).
	
	\bibitem{wrkim2011} Kim, W.\ \emph{et al}. Tunneling-Induced Spectral Broadening of a Single Atom in a Three-Dimensional Optical Lattice. Nano Lett. {\bf 11}, 729-733 (2011).
	
	\bibitem{10.1038/ncomms7977} Knebel, J., Weber, M.\ F., Kr\"{u}ger, T.\ \& Frey, E. Evolutionary games of condensates in coupled birth–death processes. Nat. Commun. {\bf 6}, 6977 (2015)
	
	\bibitem{Hong:06} Hong,\ H. -G.\ \emph{et al}. Spectral line-shape measurement of an extremely weak amplitude-fluctuating light source by photon-counting-based second-order correlation spectroscopy. Opt.\ Lett.\ {\bf 31}, 3182-3184 (2006).
	
	\bibitem{PhysRevLett.107.243002} Chalony, M.\ \emph{et al}. Doppler Cooling to the Quantum Limit. Phys.\ Rev.\ Lett.\ {\bf 107}, 243002
	(2011).
	
	\bibitem{PhysRevA.54.2275} Walhout, M., Sterr, U.\ \& Rolston, S.\ L. Magnetic inhibition of polarization-gradient laser cooling in ${\mathrm{\ensuremath{\sigma}}}^{+}-{\mathrm{\ensuremath{\sigma}}}^{\mathrm{\ensuremath{-}}}$ optical molasses. Phys.\ Rev.\ A.\ {\bf 54}, 2275-2279 (1996).
	
	\bibitem{jrjr} Kim, J.-R.\ \emph{Spectroscopic measurement of sub-Doppler cooling with two color $\sigma^{+}-\sigma^{-}$ laser configuration}, Ph.D. thesis, Seoul National University (2017).
	
	\bibitem{Castin:89} Castin, Y., Wallis, H.\ \& Dalibard, J. Limit of Doppler cooling. J.\ Opt.\ Soc.\ Am.\ B.\ {\bf 6}, 2046-2057 (1989).
	
	\bibitem{Stenholm:85} Stenholm, S. Dynamics of trapped particle cooling in the Lamb--Dicke limit. J.\ Opt.\ Soc.\ Am.\ B.\ {\bf 2}, 1743-1750 (1985).
	
	\bibitem{Molmer:93} M{\o}lmer, K., Castin, Y.\ \& Dalibard, J. Monte Carlo wave-function method in quantum optics. J.\ Opt.\ Soc.\ Am.\ B.\ {\bf 10}, 524-538 (1993).
	
	\bibitem{GE201287} Ge, H., Qian, M.\ \& Qian, H. Stochastic theory of nonequilibrium steady states. Part II: Applications in chemical biophysics. Phys.\ Rep.\ {\bf 510}, 87-118 (2012).
	
	\bibitem{1742-5468-2007-07-P07012} Zia, R.\ K.\ P.\ \& Schmittmann, B. Probability currents as principal characteristics in the statistical mechanics of non-equilibrium steady states. J.\ Stat.\ Mech.\ Theor.\ Exp.\ (2007) P07012.
	
	\bibitem{1742-5468-2007-07-P07023} Derrida, B. Non-equilibrium steady states: fluctuations and large deviations of the density and of the current. J.\ Stat.\ Mech.\ Theor.\ Exp.\ (2007) P07023.
	
	\bibitem{PhysRevA.76.013402} Choi, Y.\ \emph{et al}. Direct measurement of loading and loss rates in a magneto-optical trap with atom-number feedback. Phys.\ Rev.\ A {\bf 76}, 013402 (2007).
	
	\bibitem{PhysRevE.95.012113} Daga, B., Mondal, S., Chandra, A.\ K., Banerjee, T.\ \& Basu, A. Nonequilibrium steady states in a closed inhomogeneous asymmetric exclusion process with generic particle nonconservation. Phys.\ Rev.\ E {\bf 95}, 012113 (2017).
	
	\bibitem{Dalibard:89} Dalibard, J.\ \& Cohen-Tannoudji, C. Laser cooling below the Doppler limit by polarization gradients: simple theoretical models. J.\ Opt.\ Soc.\ Am.\ B {\bf 6}, 2023-2045 (1989).
	
	\bibitem{atomics} Foot, C.\ J.\ \emph{Atomic physics} (Oxford University Press, Oxford, 2008).
	
	\bibitem{doi:10.1119/1.4833557} Dimeo, R.\ M. Wave packet scattering from time-varying potential barriers in one dimension. Am.\ J.\ Phys.\ {\bf 82}, 142-152 (2014).
	
	\bibitem{QM} Merzbacher, E. \emph{Quantum mechanics} (John Wiley \& Sons, New Jersey, 1998).
	
\end{thebibliography}
\end{document}